\documentclass[9pt,twocolumn,twoside]{article}
\usepackage{color}
\usepackage[colorinlistoftodos]{todonotes}
\usepackage[normalem]{ulem}
\usepackage[affil-it]{authblk}
\usepackage[square,numbers,compress]{natbib}
\title{Spatially Resolved Fourier Transform Spectroscopy\\ in the Extreme Ultraviolet}

\author{G.S.M. Jansen}
\author{D. Rudolf}
\author{L. Freisem}
\author{K.S.E. Eikema}
\author{S. Witte}

\affil{\small Advanced  Research  Center  for  Nanolithography  (ARCNL), Science  Park  110,  1098  XG  Amsterdam,  The  Netherlands\newline 
Dept.  of  Physics  and  Astronomy,  Vrije  Universiteit, De  Boelelaan  1081,  1081  HV  Amsterdam,  The  Netherlands \newline\newline
email: g.s.m.jansen@vu.nl, witte@arcnl.nl}

\date{}

\begin{document}

\twocolumn[
\begin{@twocolumnfalse}
\maketitle

%\ociscodes{670.7480, 340.7480, 300.6560, 300.6540, 190.0190, 190.2620}

%\doi{\url{http://dx.doi.org/10.1364/optica.XX.XXXXXX}}

\begin{abstract}
\noindent Coherent extreme ultraviolet (XUV) radiation produced by table-top high-harmonic generation (HHG) sources provides a wealth of possibilities in research areas ranging from attosecond physics to high resolution coherent diffractive imaging. However, it remains challenging to fully exploit the coherence of such sources for interferometry and Fourier transform spectroscopy (FTS).  This is due to the need for a measurement system that is stable at the level of a wavelength fraction, yet allowing a controlled scanning of time delays.
Here we demonstrate XUV interferometry and FTS in the 17-55~nm wavelength range using an ultrastable common-path interferometer suitable for high-intensity laser pulses that drive the HHG process. This approach enables the generation of fully coherent XUV pulse pairs with sub-attosecond timing variation, tunable time delay and a clean Gaussian spatial mode profile. We demonstrate the capabilities of our XUV interferometer by performing spatially resolved FTS on a thin film composed of titanium and silicon nitride.
\end{abstract}

\end{@twocolumnfalse}
]

%\setboolean{displaycopyright}{true}

\begin{figure*}[thb!]
\centering
\includegraphics[width=1\linewidth]{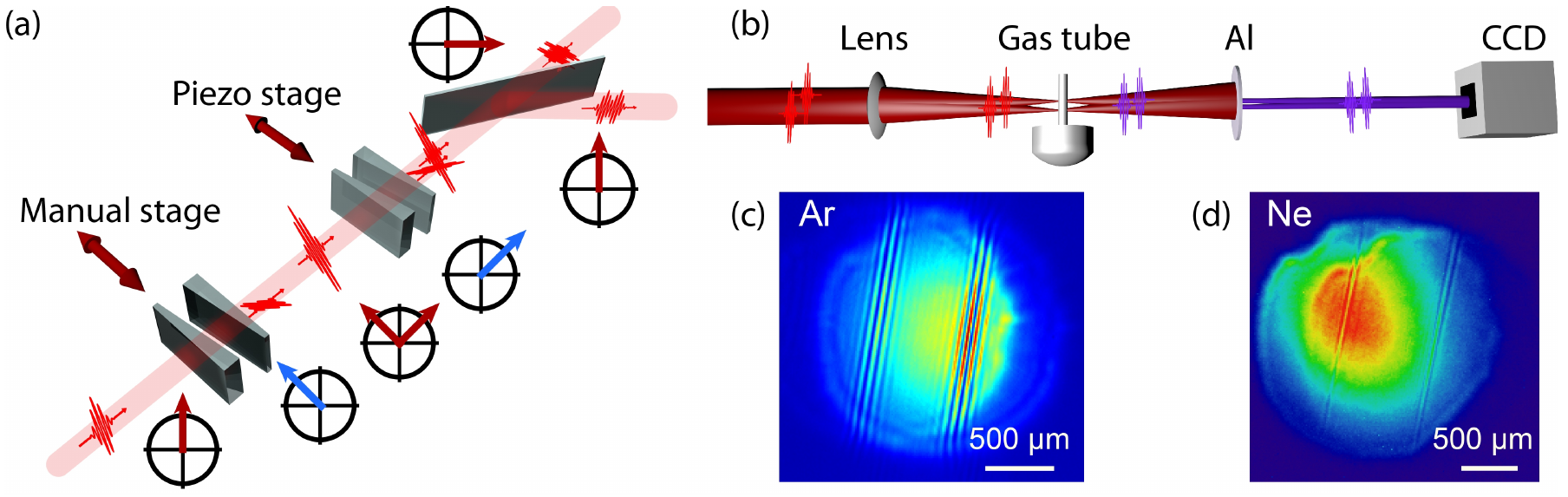}
\caption{\textbf{a:} Schematic overview of the common-path, birefringent wedge-based interferometer. The polarization diagrams depict the polarization (in red) at various positions in the interferometer. The fast axes of the birefringent wedges are indicated using blue arrows. A manual translation stage is used to control the total optical path length through the first wedge pair, while a piezo-driven stage controls the position of the second pair.  
\textbf{b:} Schematic overview of the setup for high-harmonic generation. A lens ($f=25$~cm) focuses the input pulses in the gas jet. A thin tube (inner diameter 1.4~mm) is used to guide the gas from a pulsed nozzle to the interaction region 8~mm behind the nozzle. A 1~mm aperture (not shown) blocks the near-infrared light just before the aluminum film (Al), while transmitting the high harmonics. 
\textbf{c}, \textbf{d:} Spatial interference patterns for high harmonics generated in argon and neon, respectively. 
%A movie displaying the full time-delay scan from which \textbf{c} is taken is shown in Visualization 1.
}
\label{fig:full_setup}
\end{figure*}

A well-known feature of high-harmonic generation (HHG) is broadband spectra in the XUV and soft X-ray regions \cite{krausz_attosecond_2009,popmintchev_attosecond_2010,lewenstein_theory_1994}. This radiation is typically emitted in a train of attosecond pulses with excellent spatial and temporal coherence, as shown in various interferometric and spectroscopic measurements \cite{zerne_phase-locked_1997,bellini_temporal_1998,salieres_frequency-domain_1999,Papadogiannis1999,descamps_extreme_2000,kovacev_extreme_2005,nabekawa_interferometry_2009,kandula_extreme_2010,Meng2016}. As a result, interferometry with high harmonics found important applications in e.g. Molecular Orbital Tomography \cite{Bertrand}, in wavefront reconstruction \cite{austin_lateral_2011} and electric field characterization \cite{Wyatt} of high harmonics. 
Recently, interferometry with high harmonics provided added value to coherent diffractive imaging (CDI) \cite{Witte_2014, Meng} using the full high harmonics bandwidth and photon flux. However, in the extreme ultraviolet (XUV) spectral range, interferometry and Fourier transform spectroscopy (FTS) are challenging due to the high stability requirements of the interferometer itself. 
Two main types of HHG interferometers have been devised. In one scheme, the near-infrared fundamental driving pulse is split into two phase-locked pulses with an adjustable time delay, and this pulse pair is subsequently used for HHG~\cite{bellini_temporal_1998, Papadogiannis1999, descamps_extreme_2000,  kovacev_extreme_2005}. Although this method has been successfully used it is typically limited by the stability of the optical interferometer. The other scheme is based on wavefront division, whereby one HHG beam is divided into two phase-locked sources by a piezo-mounted split mirror. This configuration allows more stable interferometry \cite{nabekawa_interferometry_2009,klisnick_experimental_2006,de_oliveira_high-resolution_2011,terschlusen_measuring_2014,Chen10062014}, but results in two beams with different spatial profiles and strong diffraction effects due to the hard edge of the split mirror. Wavefront division interferometry is also less flexible when one would like to change the intensity ratio between the two beams.

In this letter we present XUV interferometry using a novel ultrastable common-path interferometer with a timing stability better than 0.8~attoseconds (as) between two phase-locked high-harmonic sources. We characterize high-harmonic spectra from argon and and neon, and perform Fourier transform spectroscopy on a thin titanium/silicon nitride bilayer. Our approach combines the flexibility of near infrared (NIR) interferometers with the stability of common-path techniques. Because no XUV optics are involved the only bandwidth limitation is the phase-matching bandwidth of the HHG process, which can span more than an octave~\cite{popmintchev_attosecond_2010}.

\begin{figure}[!tb]
\centering
\includegraphics[width=\linewidth]{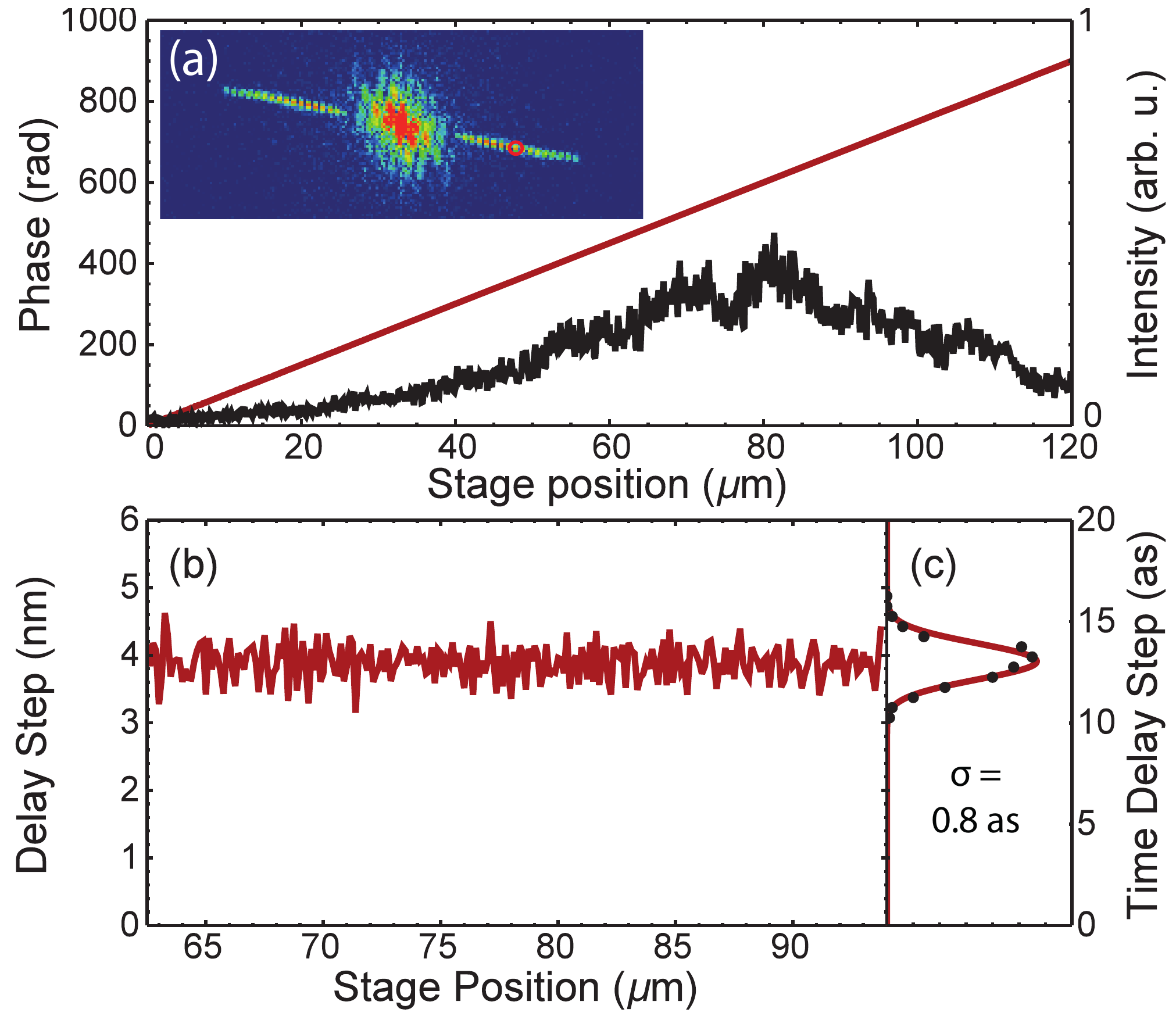}
\caption{\textbf{a:} Inset: spatial Fourier transform of the interference pattern of HHG in neon (enlarged version in Fig.~\ref{fig:sfft_si}). Phase (red) and amplitude (black) of the 33$^{\textrm{rd}}$ harmonic (red circle in the inset) as a function of piezo stage position. \textbf{b:}~Consecutive step sizes retrieved from the measured phase. \textbf{c:}~Step size distribution averaged over all harmonics has a standard deviation of 0.25~nm or 0.8~as.}
\label{fig:sfft}
\end{figure}

% Moving to the setup
In our experiments we use a Ti:Sapphire-seeded noncollinear optical parametric chirped-pulse amplifier (NOPCPA) \cite{witte_ultrafast_2012}. The pump laser for this system delivers 80~mJ, 532~nm pulses with a duration of 64~ps at 300~Hz \cite{noom_high-energy_2013}. The output of the NOPCPA is compressed to pulses of approximately 20~fs duration and an energy of 5~mJ.These pulses are then directed into our common-path interferometer, which is schematically displayed in Fig. \ref{fig:full_setup}(a). The phase-locked pulse pair is produced in birefringent wedges, where the optical axis is oriented at 45$^{\circ}$ with respect to the input polarization~\cite{brida_phase-locked_2012,oriana_FT_vis_2016}. This pair of wedges splits the input pulse into two parts of equal intensity but with orthogonal polarization, and introduces a delay of several picoseconds between the two electric field components. The second pair of wedges has its optical axis perpendicular to the first pair, thus providing an opposite delay compared to the first pair of wedges. The exact remaining delay can be controlled by moving one of the wedges perpendicular to the beam. The final delay depends linearly on the wedge displacement and can be tuned with sub-attosecond resolution.
Behind the second pair of wedges we use an ultrabroadband thin-film polarizer to project the pulses onto the same polarization axis. The resulting pulse pair is vertically polarized and contains up to 1~mJ per pulse. The intensity ratio between the beams can be controlled by tuning the input polarization state with a half-wave plate.

Behind the interferometer, the pulses are focused into a gas jet for HHG, as shown in Fig. \ref{fig:full_setup}(b). 
By tilting the last wedge of the interferometer, we ensure that the focal spots of the two pulses are separated by 280~$\mu$m in the focal plane, which is several times the focused beam diameter (60~$\mu$m $1/e^2$-diameter). This separation ensures that the two pulses generate high harmonics independently, as any partial overlap could lead to delay-dependent interference effects between the pulses, which in turn would affect ionization and phase matching. Even for this large separation, however,
the remaining field strength between the beams still leads to small modulations of the interferogram at the fundamental frequency.
Behind the HHG chamber we use an aluminum filter in combination with a 1~mm diameter aperture to separate the XUV from the driving NIR radiation. The XUV beams overlap and interfere in the far field, where we use an XUV-sensitive CCD camera (Andor iKon-L) to record the interference patterns. Examples of the measured interference patterns are given in Figs.~\ref{fig:full_setup}(c,d). The beams interfere at an apex angle of 0.4~mrad, leading to multiple zones of straight fringes with a high visibility. The spatial mode profile of the individual HHG beams has a smooth Gaussian shape. Some slight diffraction features can be observed at the edges caused by the aperture in the beam path. For more details about the interferometer and HHG see the Supplementary Information.

A \textit{spatial} Fourier transform of an interference pattern at a fixed delay directly yields an HHG spectrum~\cite{padget_staticFT_1995}, as shown in the inset in Fig.~\ref{fig:sfft}(a), which is possible because of the near-diffraction-limited beam profile of the individual HHG beams. 
Although this spectrum is limited in resolution due to the small angle, the individual harmonics are still clearly resolved. This spatial transform is useful as a single-shot diagnostic of the high harmonics spectrum. A further advantage is that these individual harmonic peaks contain information on the delay between the pulses. The change in delay between two images can be calibrated from these measurements by evaluating the phase delay of the individual harmonic peaks as a function of time delay. 
The extracted phase for a single harmonic vs. stage position is shown in Fig.~\ref{fig:sfft}(a), and confirms the scan linearity of our interferometer. In addition, the intensity of the particularly selected harmonic extracted from the interferogram is also plotted. By calculating the phase delay per stage step and dividing by the central angular frequency of the harmonic, we obtain the time delay per step (Fig. \ref{fig:sfft}(b)), as well as a measurement of the interferometer stability.  
Analyzing the data for the high-amplitude range of the scan between 60~$\mu$m and 95~$\mu$m, we find an upper limit to the timing stability of 0.8~attoseconds (standard deviation) or, equivalently, 0.25~nm optical path length stability (Fig.~\ref{fig:sfft}(c)), with a measurement accuracy limited by the precision of the phase determination in the spatial Fourier transform. It is worth noting that this upper limit includes the effects of possible differential phase shifts between the two separated HHG zones. Small intensity variations could in principle lead to phase shifts between the harmonic beams if the driving pulses are not identical \cite{bellini_temporal_1998}, but the present measurement shows that this effect does not limit the applicability of HHG-based FTS for the current spectral range. 
With the achieved stability, simulations show that FTS is feasible even at wavelengths well below 10~nm~\cite{Witte_2014}.

\begin{figure}[!tb]
\centering
\includegraphics[width=\linewidth]{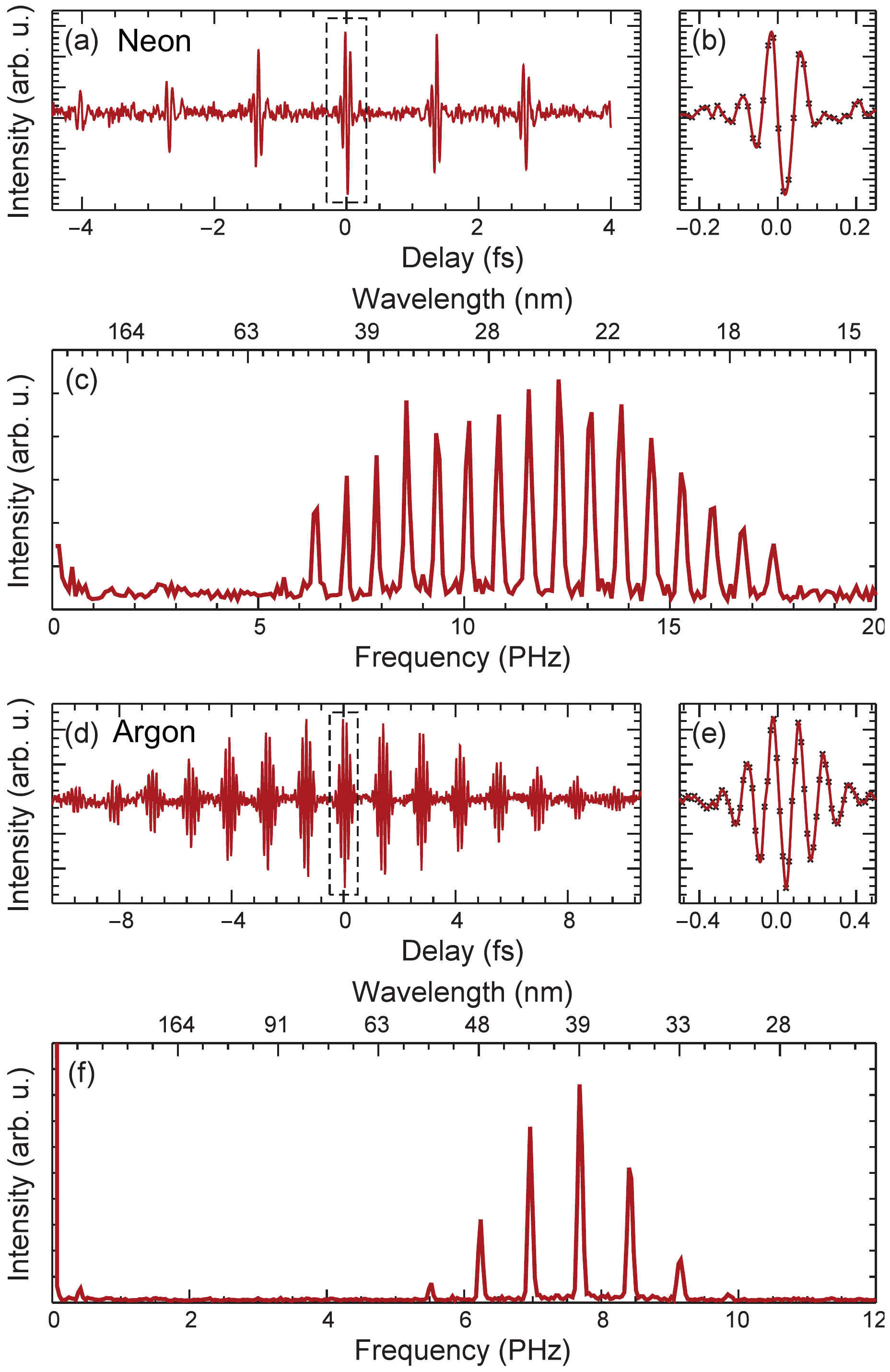}
\caption{Interferograms from a single pixel for high harmonics produced in neon (\textbf{a}) and argon (\textbf{d}) with a magnified view of the central part (\textbf{b},\textbf{e}).  
A Fourier transform of the interferograms yields the neon (\textbf{c}) and argon (\textbf{f}) spectra.}
\label{fig:spectra}
\end{figure}

By scanning the time delay between the pulses, an accurate measurement of the HHG spectrum can be obtained using FTS. We recorded such Fourier scans of high harmonics generated in argon and neon. In FTS, the step size should be less than half of the shortest wavelength in the source spectrum, and the obtained spectral resolution is determined by the length of the scan. We typically record up to a thousand images with a step size of 15~as. To remove the influence of intensity fluctuations in the recorded interference patterns (caused by laser power or beam pointing variations) and subsequently improve the signal-to-noise ratio, we normalized the spatial interference patterns to a selected local area in the beam. 
Finally, a spectrum for every pixel is acquired by Fourier transforming the measured data along the time axis. Fig.~\ref{fig:spectra} shows temporal interferograms for single pixels and the corresponding HHG spectra. As the interference pattern will only show those spectral components that are present in both beams, both driving pulses should produce identical HHG spectra, which has been confirmed using a
grazing-incidence grating XUV spectrometer. The interferograms in Figs.~\ref{fig:spectra}(a,b) and Figs.~\ref{fig:spectra}(d,e)  contain two clear timescales, corresponding to the autocorrelation widths of the individual attosecond pulses and the coherence time of the attosecond pulse train, respectively.
For the autocorrelation widths, we obtain (396 $\pm$ 10) as (Ar) and (115 $\pm$ 6) as (Ne), which reflects the about 3.5 times broader spectrum of HHG from Ne compared to Ar. On the other hand, the coherence times are (11.9 $\pm$ 0.3) fs (Ar) and (6.0 $\pm$ 0.6) fs (Ne) indicating that high harmonics in Ne are produced by the most intense temporal part of the laser pulse. The measured autocorrelation width for Ne is slightly increased by the limited transmission spectrum of the Al filter.
In contrast to a grating spectrometer, the frequency resolution in FTS is constant over the full spectrum,
which amounts to 80~THz for the example shown in Fig.~\ref{fig:spectra}(c). Expressed as a fraction of the wavelength, this means a resolution of 1 in 200 for the highest harmonic orders versus 1 in 100 for the longer wavelengths. In our current interferometer we can scan up to 50~fs in time delay, which corresponds to a potential resolution of 1 in 900 for the highest harmonic orders. This scan range can easily be extended by increasing the travel range of the piezo stage and the size of the wedges. This flexibility of the spectral range and resolution are clear advantages of FTS as a spectroscopic technique. 

\begin{figure}[tb]
\centering
\includegraphics[width=\linewidth]{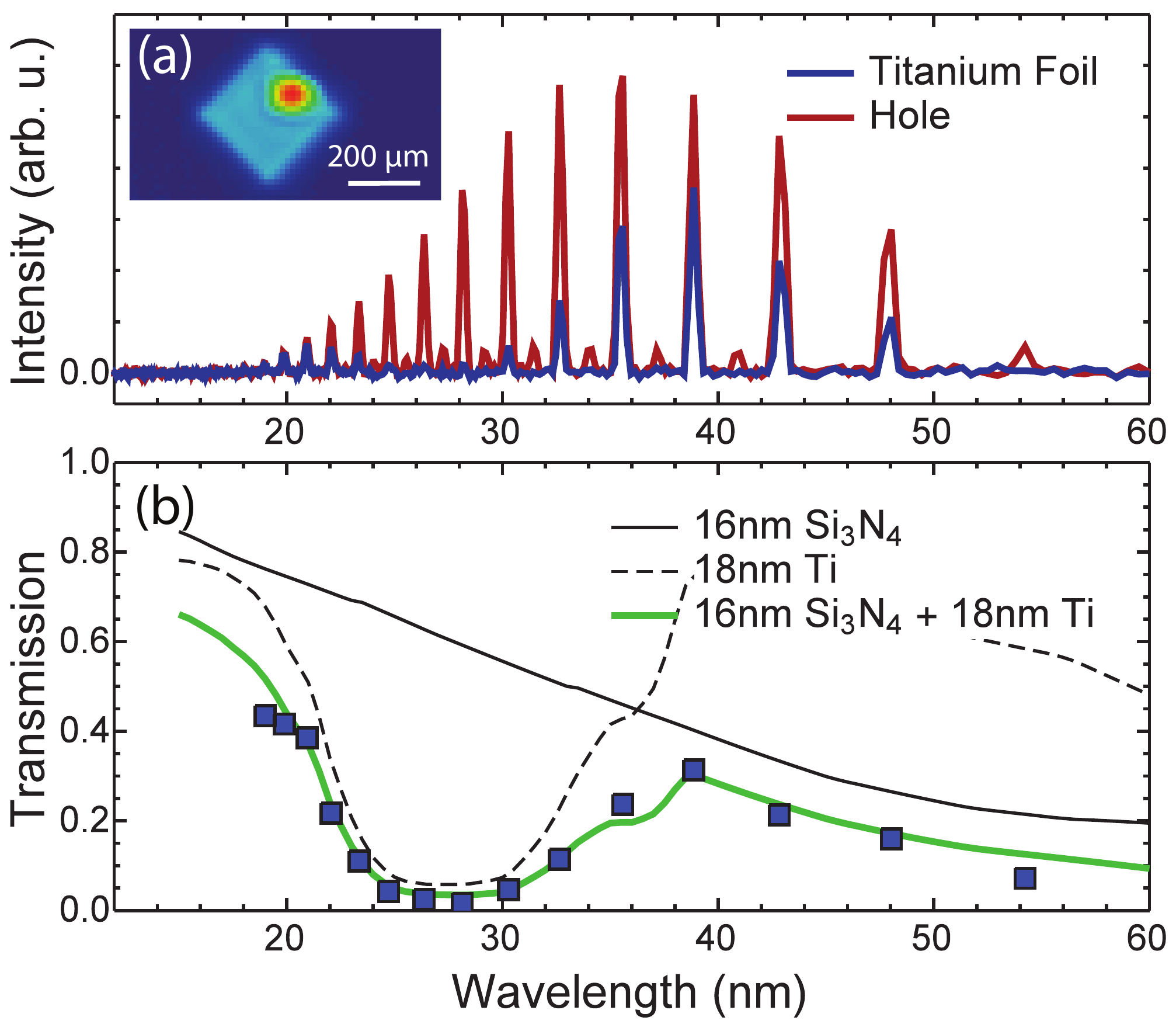}
\caption{Spatially dependent spectrum of a thin titanium/silicon nitride bilayer with an aperture. \textbf{a:}~Retrieved spectra for both the aperture and the bilayer from a single Fourier scan. In the bilayer spectrum the Ti absorption band is clearly visible. The inset shows a transmission image of the bilayer with the hole at the top right side. \textbf{b:}~The spectral transmission of the bilayer (blue squares) matches with the expected transmission of 18~nm Ti on 16~nm silicon nitride \cite{cxro}.}
\label{fig:foil}
\end{figure}

In addition to the spectral and spatial characterization of the HHG beam itself, our method can also be applied to perform spectroscopy on spatially complex samples. We explore this option by measuring the transmission spectrum of a 20~nm thin titanium film grown by electron beam evaporation on a 15~nm thin silicon nitride membrane. The membrane contains a 100~µm diameter aperture near one side (Fig.~\ref{fig:foil}(a)). With spatially resolved FTS we can simultaneously measure the XUV spectrum transmitted through the bilayer and through the aperture, a measurement that would be challenging to perform with grating-based spectrometers. We used the spectrum transmitted through the aperture as a reference for a direct determination of the relative absorption spectrum of the bilayer. 
The titanium sample was positioned in the XUV beam ensuring that both XUV pulses were overlapping on the sample. A typical transmission spectrum is given in Fig.~\ref{fig:foil}(a). A single Fourier scan yields spectra for the transmission of both the aperture and the titanium thin film (Fig.~\ref{fig:foil}(b)). The spectrum transmitted by the bilayer shows a clear dip around 25~nm matching with the corresponding absorption band of titanium. Comparing the strength of the harmonics in both spectra yields the spectral transmission of the combined titanium and silicon nitride layer, as shown in Fig.~\ref{fig:foil}(b). The measured transmission matches with the expected transmission of an 18~nm titanium layer on top of a 16~nm layer of silicon nitride \cite{cxro}. Potentially, there could be a thin titanium oxide layer on top of the titanium film, but given the small difference in XUV absorption this distinction cannot be made.

In summary, we have demonstrated XUV interferometry and Fourier transform spectroscopy without the need for XUV optics. Using a birefringence-based common-path interferometer, we achieve sub-attosecond timing stability and a high spectral resolution. With two HHG pulses from neon gas, we measure more than octave-spanning spectra down to 17~nm wavelength
and find that the HHG process adds less than 0.8~as relative timing jitter under our experimental conditions. 
We demonstrate that our FTS method can be used to measure the absorption spectrum of a spatially inhomogeneous thin film sample. For future experiments it is particularly promising to combine this method with CDI on nanostructures composed of multiple materials.

\section*{Funding and Acknowledgements}
\vspace{\baselineskip}
\noindent
%\textbf{Funding and Acknowledgement}
The project has received funding from the European Research Council (ERC) (ERC-StG 637476) and the Foundation for Fundamental Research on Matter (FOM), which is part of the Netherlands Organisation for Scientific Research (NWO). 

%\section*{Acknowledgments}
%\vspace{\baselineskip}
%\noindent
%\textbf{Acknowledgement.}
%We thank F. Melsheimer and Prof. L. Juschkin for the use of their EUV spectrometer, and R. Kortekaas for technical assistance.

\vspace{\baselineskip}
\noindent 
Supplementary information follows at the end of this manuscript. 

% Bibliography 
%\bibliography{XUVinterferometry}

\clearpage

\twocolumn[

\begin{@twocolumnfalse}
\centering \section*{SUPPLEMENTARY INFORMATION}
\vspace{2cm}
\end{@twocolumnfalse}
]

\section*{Birefringent wedge-based interferometer}
In the interferometer, a piezo linear stage (Physik Instrumente GmbH, model number P-625.1CD) is used to displace one of the birefringent wedges. The stage has a travel range of 500~$\mu$m, a resolution of 1.4~nm and a repeatability of approximately 5~nm. 

The final delay introduced by the wedge-based interferometer depends linearly on the wedge displacement $\Delta x$ and is given by $\Delta t = \left(\Delta n \Delta x / c\right) \tan \varphi$,
where $\Delta n = n_e - n_o$ is the difference in refractive index for the extraordinary and ordinary polarization states, $c$ is the speed of light in vacuum and $\varphi = 15^{\circ}$ is the apex angle of the wedges. We use $\alpha$-BBO crystals in the interferometer because of their strong birefringence ($\Delta n = 0.11$ at 800~nm) and low nonlinear susceptibility. Behind the second pair of wedges we use an ultrabroadband thin-film polarizer to project the pulses onto the same polarization axis.  
The apex angle of the birefringent wedges was chosen to be 15 degrees, leading to a full scan range of 50~fs. Based on the specifications of the piezo stage and the refractive indices of $\alpha$-BBO, the resolution and repeatability of the interferometer are 0.14 as and 0.5 as, respectively. 
The $\alpha$-BBO wedges are mounted using KM100CL rectangular mounts (Thorlabs) at a height of 8 centimeters above the baseplate. This baseplate is bolted directly to the optical table. Care was taken to keep the optical components on low and stable posts to minimize potential vibrations. 

To minimize vibrations caused by the vacuum system influencing high-harmonic generation (HHG), we employ low-vibration turbomolecular pumps (Pfeiffer HiPace 700), while the scroll pump (Edwards XDS10) used for backing the turbos is connected using flexible bellows wrapped in vibration-damping foam.

\begin{figure*}[!ht]
\centering
\includegraphics*[width=\linewidth]{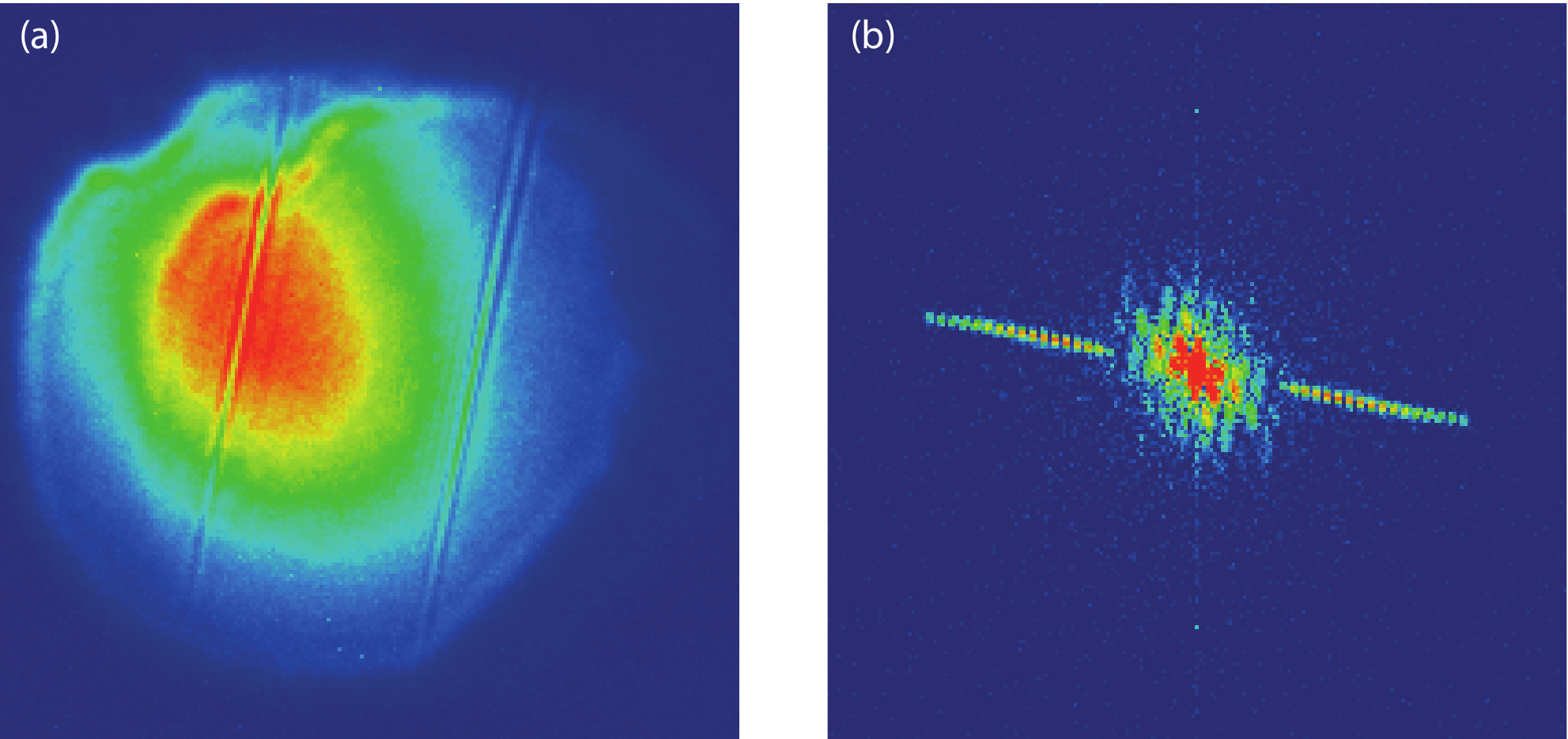}
\caption{\textbf{a:}~Typical spatial interference pattern as recorded for a fixed time delay using high harmonics generated in Neon gas.  
\textbf{b:}~Amplitude of the two-dimensional Fourier transform of the example interference pattern in (a).}
\label{fig:sfft_si}
\end{figure*}

\section*{High-harmonic generation geometry}

Behind the interferometer, a lens ($f=25$~cm) focuses the pulses into a gas jet for high-harmonic generation. The gas is supplied by a pulsed piezo valve (developed by M.H.M. Janssen, Vrije Universiteit Amsterdam) and guided through a thin tube towards the interaction region with near-infrared pulses.  
To ensure that both near-infrared pulses generate XUV independently, we separate the foci by 280~$\mu$m. In the chosen geometry this means that one pulse is focused slightly closer to the gas nozzle than the other pulse. 
We focus the pulse pair into the gas jet at a fairly large distance of about 8~mm with respect to the nozzle. Simulations of supersonic gas expansion show that the pressure difference between the two focal spots at this nozzle distance is sufficiently small to ensure similar phase matching conditions between the two beams. 
In addition, for gas confinement we use a thin stainless steel tube with an inner diameter of 1.4~mm to guide the gas from the nozzle to the interaction region.  
The near-infrared laser pulses are focused through a 0.5~mm aperture drilled perpendicular to the tube axis. 
To confirm that the phase matching conditions are similar for both pulses, we used a grazing incidence XUV grating spectrometer to measure the HHG spectrum of the individual pulses. The XUV spectrometer, equipped with a 600 lines/mm grating, is designed for a spectral range between 5 nm  and 40 nm and was operated at a resolution of $\lambda/\Delta \lambda=$ 250 at 20 nm wavelength. As for single high harmonic beams, we observed the spectra to be identical. Furthermore, the HHG spectrum observed for both simultaneously present beams did not change, confirming that the pulses did not influence each other significantly. 

While the chosen geometry worked well and was experimentally convenient to realize, a geometry where the two pulses are displaced orthogonally with respect to the gas jet direction can potentially lead to slightly better phase matching conditions. Such a geometry, however, does require precise knowledge of the transverse pressure profile in the gas jet, and a sufficiently wide jet compared to the focal spot separation.

\section*{Single-shot Fourier Transform interferometry}

The generated XUV beams overlap with an apex angle of approximately 0.4 milliradians. Therefore, for each harmonic a fringe pattern in close analogy to the Young's double slit experiment is formed. The period of these fringe patterns is proportional to the wavelength of the respective harmonic, while the phase of the fringe is determined by the delay between the two pulses. The fringe patterns for all harmonics then add up coherently to form the interference pattern as measured on the camera.  In Fig.~\ref{fig:sfft_si}(a), the plane spanned by the k-vectors of the two beams (interference plane) is rotated with respect to the horizontal plane, which results in a slightly tilted interference pattern with respect to the vertical axis of the camera image. 

The raw camera images can be decomposed into the contributions of the individual harmonics by a two-dimensional Fourier transform of the raw image. For a fixed angle between the beams, every harmonic wavelength corresponds to a specific spatial frequency in the Fourier transform image. Therefore, the individual harmonics appear spatially separated in the Fourier transform data. The center of the image corresponds to zero spatial frequency, and increasing spatial frequencies are positioned radially outward. 
This yields a set of peaks corresponding to the individual harmonics lying on a line perpendicular to the original fringe direction, as shown in Fig.~\ref{fig:sfft_si}(b). The phase of the individual peaks in this spectrum directly yields the phase of the corresponding monochromatic fringe pattern. 
Due to the symmetry of the Fourier transform for real data, each harmonic gives rise to two peaks, at both positive and negative spatial frequency.
Comparing the phase between two measurements yields a value that is proportional to the change in delay between these measurements. If the wavelength of the harmonic is known, this can be used to extract the exact change in delay.

\end{document}